%% file: A_Nested_Amplitude_Amplification_protocol_for_the_binary_knapsack_problem.tex
\documentclass[10pt,conference]{IEEEtran}
\IEEEoverridecommandlockouts

\usepackage{cite}
\usepackage{amsmath,amssymb,amsfonts}
\usepackage{graphicx}
\usepackage{textcomp}
\usepackage{xcolor}
\usepackage{algpseudocode}
\usepackage{algorithm}
\usepackage{amsmath}

\DeclareMathOperator*{\argmin}{argmin}
\usepackage{braket} 
\usepackage{lipsum}
\usepackage{quantikz}
\usepackage[hidelinks]{hyperref}

\def\BibTeX{{\rm B\kern-.05em{\sc i\kern-.025em b}\kern-.08em
    T\kern-.1667em\lower.7ex\hbox{E}\kern-.125emX}}
    
\begin{document}

\title{A Nested Amplitude Amplification Protocol for the Binary Knapsack Problem}

\author{\IEEEauthorblockN{1\textsuperscript{st} Laurin Demmler}
\IEEEauthorblockA{\textit{Infineon Technologies AG} \\
Munich, Germany \\
laurin.demmler2@infineon.com}
\and
\IEEEauthorblockN{2\textsuperscript{nd} Maximilian Hess}
\IEEEauthorblockA{\textit{Infineon Technologies AG} \\
Munich, Germany}}

\maketitle

\begin{abstract}
\input{Abstract.tex}
\end{abstract}


\section{Introduction}\label{Introduction}
\input{Introduction.tex}
\section{Background}\label{background}
\subsection{The Binary Knapsack problem}\label{knapsack}
\input{Knapsack}
\subsection{The Quantum Tree Generator}\label{QTG}
\input{QTG}

\subsection{Grover Adaptive Search}\label{GAS}
\input{GAS}

\section{Nested Amplitude Amplification}
\input{Nested_AA}
\section{Classical Simulation of Amplitude Amplification Circuits}
\label{classical_tracking}
\input{ClassicalTracking.tex}

\section{Results}
\label{results}
\input{results.tex}

\section{Outlook}
\input{outlook.tex}
\section*{Acknowledgment}
This work was supported by the BMFTR funded project QuSol. We thank Prof. Christian Mendl for the insightful discussions.

\section*{Data and Code Availability}
The code and depicted data can be accessed at \url{https://github.com/LaurinDemmler/A_Nested_AA_protocol_binary_knapsack}.

\bibliographystyle{IEEEtran}
\bibliography{refs}
\begin{appendices}
\section{Oracle Construction}\label{oracledesign}
\input{OracleConstruction}
\section{Results for Weakly Correlated Instances}\label{weak_results}
\input{Weakly_Correlated_Results}

\end{appendices}

\end{document}

%% file: Abstract.tex
Amplitude Amplification offers a provable speedup for search problems, which is leveraged in combinatorial optimization by Grover Adaptive Search (GAS). The protocol demands deep circuits that are challenging with regards to NISQ capabilities. We propose a nested Amplitude Amplification protocol for the binary knapsack problem that splits the decision tree at a tunable depth, performing a partial amplification on the first variables before executing a global GAS on the full search space. The partial amplification is implemented by an Inner Iteration Finder that selects the rotation count maximizing marked-subspace amplitude. The resulting biased superposition serves as the initial state for the outer Amplitude Amplification. Using the Quantum Tree Generator for feasible-state preparation and an efficient classical amplitude-tracking scheme, we simulate the protocol on knapsack instances of sizes intractable by statevector simulation. Our results show that the nested approach reduces the cost of improving an incumbent solution compared to baseline GAS, particularly for a specific subset of knapsack instances. As combinatorial problems in domains such as semiconductor supply-chain planning grow in scale, methods that reduce circuit cost are an important step toward eventual quantum advantage for such applications.  

%% file: Introduction.tex
The knapsack problem and its variants are an essential cornerstone of industrially relevant combinatorial optimization \cite{knapsack_industry1,knapsack_industry2}. In particular, large-scale and increasingly automated global supply chains, such as those found in the semiconductor industry, rely on knapsack-type formulations for resource allocation and capacity planning \cite{global_fab,knapsack_infineon}. While efficient classical algorithms exist, they can be computationally intractable for industrial sized instances \cite{WhereAreTheHardKnapsackInstances,knapsack_classical}. Motivated by the restrictions of classical combinatorial optimization on industry-relevant problems, the field of quantum optimization problems has emerged as a promising candidate for quantum advantage. While quantum hardware grows in size and accuracy \cite{IBMnumberQubits,GoogleErrorCode}, there remains a significant amount of work to be done on the theoretical approaches to quantum optimization. The NISQ era has seen the emergence of many algorithms, such as QAOA \cite{QAOA}, VQE \cite{VQE} and quantum annealing \cite{Annealing} for optimization. Many of these algorithms show signs of quantum advantage for certain, constructed instances or problem classes, which however remains elusive to prove generally \cite{QAOA_advantage,Annealing_Advantage}.\\
Another approach to quantum optimization, which is likely to become increasingly relevant as hardware providers promise to leave the NISQ era behind is Amplitude Amplification (AA). In a landmark paper by Lov Grover \cite{Grover_original_paper}, Grover's algorithm was introduced as a means to perform unstructured database search on a quantum computer with quadratic speedup. Since then, modification to non-uniform superpositions \cite{AA} and quantum optimization in the form of the Grover Adaptive Search (GAS) \cite{GAS} have been proposed. The evident advantage of AA compared to the aforementioned variational approaches is manifold. The squareroot speedup over classical brute-force approaches remains in an optimization context. Furthermore, amplitudes of selected states can be tracked through the algorithm classically (and efficiently), which allows for classical simulation that transcends many of the problems faced by modern simulation methods, such as Matrix Product States (MPS) \cite{MPS1, MPS2, MPS3}. On the other side, GAS is clearly not a NISQ algorithm, both with regards to circuit depth and error sensitivity\cite{GAS_NISQ}.\\
In recent years, GAS has been frequently combined with problem specific state preparations to improve the performance of the algorithm \cite{MeinPaper, montanaro2010quantum}. The Quantum Tree Generator (QTG) \cite{QTG,QTG_on_QAOA} stands out as a particularly promising example of such state preparation, applied to the knapsack problem. It enables the generation of a biased superposition of feasible states to the problem.\\
This paper builds on the idea proposed first in \cite{NestedByGrover}. Nesting a quantum algorithm within itself, can have positive effects on its performance \cite{NestedTree,NestedQuantumWalks}. In this paper, we conceptualize and test a nested GAS protocol for the binary knapsack problem, using the QTG as a biased state preparation. We show that for certain types of problem instances, nesting a partial AA into the standard GAS protocol can lead to improved performance compared to regular AA. We further provide computational evidence on the optimal depth of the nesting.

%% file: Knapsack.tex
Studied as early as 1896 \cite{OnThePartitionOfNumbers}, the knapsack problem is a staple piece in combinatorial optimization. Given a knapsack of integer capacity $c$ and $n$ items, each described by a weight $w_i \in \mathbb{N}$ and a profit $p_i \in \mathbb{N}$, one is to pack the knapsack in such a way, that the packed profit is maximized, while the total weight does not exceed the capacity. Formally, one assigns a binary variable $x_i \in \{0,1\}$ to each item, such that
\begin{align}
    x_i &= 0 \iff \text{item $i$ is not included},\\
    x_i &= 1 \iff \text{item $i$ is included}.
\end{align}
The problem can now be expressed in the following way.
\begin{align}
    \max_{x \in \{0,1\}^n} &\sum_{i=0}^{n-1}p_ix_i, \\
    \text{subject to} &\sum_{i=0}^{n-1}w_ix_i \leq c. \label{capacity_constraint}
\end{align}
The binary knapsack problem has been studied for decades, due to different reasons. First, the problem is hard, in a sense that it is a provably NP-hard problem~\cite{Karp1972}. Secondly, while it is NP-hard, the problem displays a manageable structure, in that it maximizes a linear function over a single linear constraint. Thirdly, over the decades many knapsack instances have been found to be solvable in reasonable amount of time, spurring new quests to look for hard knapsack instances. \cite{KnapsackProblemsBook, WhereAreTheHardKnapsackInstances, KnapsackNPCompleteness, AlgorithmsForKnapsack}.
\\
For all of the reasons above, as well as the significant amount of benchmarking potential, the Quantum Optimization Community has turned towards the knapsack problem as a playground to assess the effectiveness of Quantum Algorithms \cite{QuantumKnapsack1,QuantumKnapsack2,QuantumKnapsack3}.

%% file: QTG.tex
As introduced in Sec.~\ref{Introduction}, problem specific state preparation for quantum combinatorial optimization has become commonplace. While showing promising results in boosting the performance of QAOA approaches through hot-starting \cite{QTG, QTG_on_QAOA, MeinPaper, XY-Mixer}, it can also be a powerful way of creating a favorable superposition as an input state to AA-based protocols. One particularly exciting algorithm is the so called 'Quantum Tree Generator' for the binary knapsack problem \cite{QTG}. The QTG is a way of preparing a (possibly biased) superposition of feasible states for the binary knapsack problem. 

\begin{algorithm}[H]
\caption{Quantum Tree Generator}
\label{QTG_Pseudocode}
\begin{algorithmic}[1]
\Ensure Quantum state encoding solution space
\State \textbf{Initialize:} Prepare three quantum registers:
\State \quad Item Register: $\ket{0}^{\otimes n}$
\State \quad Capacity Register $C$: $\ket{c}$ \Comment{Remaining Capacity}
\State \quad Total Profit Register $P$: $\ket{0}$ 
\For{$m = 1$ to $n$}
    \State $cH'_{w_m \leq C}(x_m)$ \Comment{Biased Hadamard on $x_m$, controlled on $w_m \leq C$}
    \State $c\mathrm{SUB}_{x_m}(C,\, w_m)$ \Comment{$C \leftarrow C - w_m$}
    \State $c\mathrm{ADD}_{x_m}(P,\, p_m)$ \Comment{$P \leftarrow P + p_m$}
\EndFor
\State \textbf{Return} $\sum_{\substack{x \in \{0,1\}^n \\ w^T x \leq c}} \alpha_x \ket{x} \ket{c- w^T x} \ket{p^T x}$
\end{algorithmic}
\end{algorithm}

Following \cite{QTG}, one can specify the QTG as Alg.~\ref{QTG_Pseudocode}, where $\alpha_x$ denotes the amplitude of state $\ket{x}$ and can be generally non-uniform. We adopt the shorthand $cU_{\text{cond}}(\text{target})$ for a gate $U$ applied to register \textit{target}, controlled on condition \textit{cond}. The idea of the QTG follows a procedure to classically sample feasible states, which is explained in the following. Given some bitstring of length $k$ and a reference solution (often the greedy solution) $x^{\text{ref}}$, one first verifies whether including item ${k+1}$ violates the remaining capacity of the knapsack. If this is not the case, one chooses $x_{k+1}=1$ with some probability (depending on whether $x^{\text{ref}}_{k+1}=1$), and $x_{k+1}=0$ otherwise. Concretely, the action of the QTG at step $k+1$ on a bitstring $x\in\{0,1\}^k$ with remaining capacity $C$ and accumulated profit $P$ is
\begin{multline}\label{QTG_branching}
    \ket{x0\dots0}\ket{C}\ket{P} \;\mapsto\; \sqrt{\mathcal{P}}\,\ket{x0\dots0}\ket{C}\ket{P} \\
    +\; \sqrt{1-\mathcal{P}}\,\ket{x10\dots0}\ket{C-w_{k+1}}\ket{P+p_{k+1}},
\end{multline}
where
\begin{equation}
    \mathcal{P} = \begin{cases}
    \dfrac{b+1}{b+2}, & \text{if } w_{k+1} \leq C \text{ and } x^{\text{ref}}_{k+1} = 0,\\[6pt]
    \dfrac{1}{b+2}, & \text{if } w_{k+1} \leq C \text{ and } x^{\text{ref}}_{k+1} = 1,\\[6pt]
    1, & \text{if } w_{k+1} > C,
    \end{cases}
\end{equation}
and the bias $b \geq 0$ determines to what degree the superposition favours the reference solution $x^{\text{ref}}$ ($b=0$ recovers the standard Hadamard). If item $k+1$ does not fit ($w_{k+1}>C$), the state is not split and the item is excluded with certainty. This approach also enables one to track amplitudes for each iteration of the QTG classically, which is described in more detail in Sec.~\ref{classical_tracking}.

%% file: GAS.tex
Grover's algorithm promises a square-root speed up over classical approaches when searching an unstructured database \cite{Grover_original_paper}. While not naively applicable to optimization problems, it can be modified in a way to produce the so called 'Grover Adaptive Search' (GAS), proposed in \cite{GAS}. GAS can be understood as a quantum-assisted sequential approximation method applied to a general optimization problem.\\
At its core, GAS is an algorithm that takes some cost function $f$, a state preparation operator $A$ and the Oracle operator $O(y)$ as input. Note that the following description assumes that $A$ prepares only feasible states. If infeasible states may be prepared, the oracle must also account for constraint violations (by e.g. penalty terms). GAS then performs a Grover search for all states that are better than the incumbent $y$ (w.r.t. $f$) using a random number of Grover iterations sampled from an exponentially increasing interval. Sampling in this manner (rather than for example from a linearly increasing interval) maintains the quadratic speed-up of the original quantum search protocol~\cite{Grover_original_paper, BBHT}. States are marked by the adaptive oracle $O(y)$, which adds a phase for all states $x$ such that $f(x) < y$.  Since this algorithm is central to this paper, the pseudo-code from \cite{GAS} is reproduced in spirit in Alg.~\ref{QSearch_pseudocode} and Alg.~\ref{GAS_pseudocode}.

\begin{algorithm}[H]
\caption{QSearch}
\label{QSearch_pseudocode}
\begin{algorithmic}[1]
\Require State preparation $A$, oracle $O(y_i)$, incumbent $x$, $y$, $k$, $\lambda$
\State Randomly select the rotation count $r$ from the set $\{0, 1, \ldots, \lceil \sqrt{k} \rceil - 1\}$;
\State Apply Grover Search with $r$ iterations, using $A$ and $O(y)$. We denote the outputs $\tilde{x}$ and $\tilde{y}$ respectively;
\If{$\tilde{y} > y$}
    \State $x = \tilde{x}$, $y = \tilde{y}$, and $k = 1$
\Else
    \State $x = x$, $y = y$, and $k = \lambda k$
\EndIf
\State \Return $x$, $y$, $k$, $r$
\end{algorithmic}
\end{algorithm}

\begin{algorithm}[H]
\caption{Grover Adaptive Search}
\label{GAS_pseudocode}
\begin{algorithmic}[1]
\Require $f : X \to \mathbb{R}$, $\lambda > 1$
\State Uniformly sample $x_1 \in X$ and set $y_1 = f(x_1)$;
\State Set $k = 1$ and $i = 1$;
\Repeat
    \State $x_{i+1}, y_{i+1}, k, r \gets \text{QSearch}(A, O(y_i), x_i, y_i, k, \lambda)$
    \State $i = i + 1$;
\Until{a termination condition is met;}
\end{algorithmic}
\end{algorithm}

It should be mentioned here that setting $A$ to
\begin{eqnarray}
    A = H^{\otimes n}
\end{eqnarray}
recovers a Grover based algorithm, whereas a general $A$ should be correctly referred to as an Amplitude Amplification based algorithm \cite{AA}. In fact later, we will define the QTG circuit to be $A$. It is in general also possible to make the state preparation operator dependent on the incumbent, $A = A_y$. The design of the oracle $O_y$ is problem specific, and is described in more detail in App.~\ref{oracledesign} for the binary knapsack problem.

%% file: Nested_AA.tex
This section is intended to describe the main idea of this paper. Ideas to implement nested versions of Tree Search \cite{NestedTree}, Quantum Walks \cite{NestedQuantumWalks} and Quantum Search on structured problems \cite{NestedByGrover} have been proposed in the past, so it is only natural to try to come up with a similar idea for GAS on knapsack problems. 
\subsection{The baseline approach}\label{baseline_approach}
The starting point is the hot-started GAS for the binary knapsack problem defined as before in Sec.~\ref{knapsack}. In total, the algorithm uses $n$ qubits to represent the decision variables, a register $c$ to represent the remaining capacity and a register $p$ to represent the current profit of the solution. The QTG from Sec.~\ref{QTG} is then applied as a state preparation $A$, followed by a GAS on the whole search space. Each of the AA steps implements the operator
\begin{eqnarray}
    \label{VanillaOPerator}
    \left(-A D A^\dagger O_{\text{global}}(y)\right)^rA,
\end{eqnarray}
where $O_{\text{global}}(y)$ is the oracle that marks all states with a profit greater than the incumbent value $y$ (defined later in \eqref{oracles}), $D$ is the diffusion operator, and $r$ is the number of Grover iterations.

Given some incumbent solution $x^\text{incumbent}$ to the binary knapsack problem, the incumbent value is
\begin{eqnarray}
    y = \sum_{i=1}^n p_i x^\text{incumbent}_i.
\end{eqnarray}
The oracle for the GAS algorithm described in Sec.~\ref{GAS} hence marks all states $x$ such that
\begin{eqnarray}
    \label{vanilla_value}
    y < \sum_{i=1}^n p_i x_i.
\end{eqnarray}
Executing the QTG Circuit before the GAS allows us to project our search space solely into the space of feasible solutions, which means that all marked states (and indeed all states that have a measurement probability greater than zero) satisfy \eqref{capacity_constraint}. 

\subsection{The nested approach}\label{nested_approach}
Performing a GAS on the whole search space requires a potentially large (and therefore expensive) number of Grover iterations. Instead, in this paper the decision tree composed of the decisions of including or excluding each item is cut at some depth $k$, where a partial capacity condition is met and a partial value criterion is amplified. 
Partial states are represented by bitstrings of length $k$. Applying the QTG circuit on the qubits of these partial states, ensures that only partially feasible states receive a non-zero amplitude. The condition at depth $k$ for some bitstring $x$ is defined as
\begin{eqnarray}
    \label{partial_capa}
    \sum_{i=1}^k w_i x_i \leq c.
\end{eqnarray}
The question that thus arises is which partial states to amplify. One of the more straight-forward approaches is to amplify all partial states that, if the remaining items were to be included, could still improve on the incumbent value $y$, i.e.
\begin{eqnarray}
    \label{partial_val}
    y - \sum_{i=k}^{n-1} p_i < \sum_{i=0}^{k-1} p_i x_i.
\end{eqnarray}
Any partial assignment up to depth $k$ which fails to satisfy condition~\eqref{partial_val} cannot be completed to a globally marked state.

As opposed to the baseline approach, the state resulting from partial AA fulfilling~\eqref{partial_capa} and~\eqref{partial_val} is subsequently not measured, but used as a biased starting state for a GAS on the whole search space. The motivation behind this approach is the possibility of significantly reducing the number of Grover iterations on the whole search space by `paying' with less expensive iterations on the space of $k$-bit strings. On the partial subspace, the AA is performed using the Inner Iteration Finder (IIF) described in Alg.~\ref{IIF_pseudocode}. Since we are ultimately not interested in measuring the state that results from the IIF, it returns the number of Grover iterations that maximize the amplitude of the marked subspace for the global GAS and the cost associated with that process. This knowledge is then used to construct the state preparation operator for the global GAS, which is a combination of the QTG on the remaining items and the inner AA.

\begin{algorithm}[H]
\caption{Inner Iteration Finder (IIF)}
\label{IIF_pseudocode}
\begin{algorithmic}[1]
\Require Knapsack instance, depth $k$, incumbent $x$, incumbent value $y$, $\lambda > 1$
\Require Number of validation samples $L$, termination $t_{\text{IIF}}$ \eqref{IIF_termination}
\State \textbf{Set} $m \gets 1$, $C_{\text{IIF}} \gets 0$
\State \textbf{Set} $y \gets y-\sum_{i=k+1}^n p_i$ and $x \gets x_{1:k}$
\State \textbf{Set} $\text{success} \gets 0$
\While{not $t_{\text{IIF}}(\text{success, L})$}
    \State \textbf{Set} $A\gets \text{QTG}_k$ \Comment{partial QTG on the first $k$ items}
    \State \textbf{Sample} $r_{\text{in}}$ from $\{0, 1, \hdots, \lceil \sqrt{m} \rceil - 1\}$
    \State \textbf{Set} $\text{success} \gets 0$ and $\text{total} \gets 0$
    \Repeat
        \State \textbf{Apply} $\left(-ADA^\dagger O_{\text{inner}}(y)\right)^{r_{\text{in}}}A$
        \State \textbf{Measure} $\tilde{x}$ and $\tilde{y}$
        \State $\text{total} \gets \text{total} + 1$
        \State $C_{\text{IIF}} \gets C_{\text{IIF}} + k(2r_{\text{in}}+1)$
        \If {$\tilde{y} > y$}
            \State $\text{success} \gets \text{success} + 1$
        \EndIf
    \Until{$\text{total} = L$ \textbf{or} $\text{success} \neq \text{total}$}
    \State \textbf{Set} $m \gets \min(\lambda m, \, 2^k)$
\EndWhile
\State \Return $r_{\text{in}}$, $C_{\text{IIF}}$
\end{algorithmic}
\end{algorithm}

The full procedure using Alg.~\ref{IIF_pseudocode} is summarized in Alg.~\ref{NestedGAS_pseudocode}. The state preparation operator $A_{global}$ replaces the standard QTG circuit used in the baseline approach described in Sec.~\ref{baseline_approach}. It is much more complex and incorporates the entire partial AA.

\begin{algorithm}[H]
\caption{Nested GAS}
\label{NestedGAS_pseudocode}
\begin{algorithmic}[1]
\Require Knapsack instance $\text{KS}$ with $n$ items, depth $k$, $\lambda > 1$
\Require Number of IIF validation samples $L$, budget $\mathcal{B}$, termination $t_{\text{GAS}}$ \eqref{t_GAS}
\State \textbf{Sample} feasible $x \in \{0,1\}^n$ 
\State \textbf{Set} $y \gets \sum_{i=1}^n p_i x_i$
\State \textbf{Set} $m \gets 1$, $C \gets 0$, $y_{\text{prev}} \gets -\infty$
\While{not $t_{\text{GAS}}(C)$}
    \If{$y \neq y_{\text{prev}}$} 
        \State $r_{\text{in}},\, C_{\text{IIF}} \gets \text{IIF}(\text{KS}, k, x, y, \lambda, L)$
        \State $C \gets C + C_{\text{IIF}}$
        \State \textbf{Set} $A \gets \text{QTG}_k$
        \State \textbf{Set} $B \gets \text{QTG}_{n-k}$ \Comment{prepare remaining items}
        \State \textbf{Set} $D \gets I-2\ket{0}\bra{0}$
        \State \textbf{Set} $A_{\text{global}} \gets B\left(-ADA^{\dagger}O_{\text{inner}}\right)^{r_{\text{in}}}A$
    \EndIf
    \State $y_{\text{prev}} \gets y$
    \State $x, y, m, r \gets \text{QSearch}(A_{\text{global}}, O_{\text{global}}(y), x, y, m, \lambda)$
    \State $C \gets C + (2r+1)(n + 2r_{\text{in}}k)$
\EndWhile
\State \Return $x$, $y$
\end{algorithmic}
\end{algorithm}
Explicitly, the oracles for the IIF and the global GAS are defined as follows:
\begin{align}
    \label{oracles}
    O_{\text{IIF}}(y)\ket{x} &= \begin{cases}
    - \ket{x}, & \text{if } \sum_{i=0}^{k-1} p_i x_i > y - \sum_{i=k}^{n-1} p_i \\
    \ket{x}, & \text{else}
    \end{cases}, \\
    O_{\text{global}}(y)\ket{x} &= \begin{cases}
    - \ket{x}, & \text{if } \sum_{i=0}^{n-1} p_i x_i > y \\
    \ket{x}, & \text{else}.
    \end{cases}
\end{align}
The specific implementation of these oracles is described in more detail in App.~\ref{oracledesign}.

\subsection{Comparing costs}
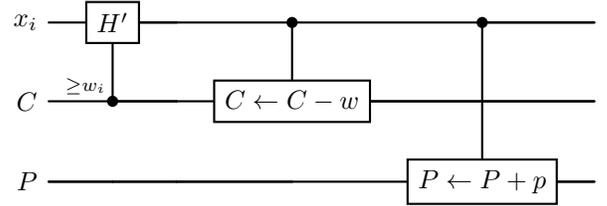
\begin{figure}[H]
\centering
\begin{quantikz}
\lstick{$x_i$} & \gate{H'} & \qw & \ctrl{1} & \ctrl{2} & \qw \\
\lstick{$C$} & \ctrl[wire style={"\geq w_i", pos=0.2}]{-1} & \qw & \gate{C \leftarrow C - w} & \qw & \qw \\
\lstick{$P$} & \qw & \qw & \qw & \gate{P \leftarrow P + p} & \qw
\end{quantikz}
\caption{The action of including or excluding a single item in the superposition is defined as a unit of cost.}\label{fig:unitofcost}
\end{figure}

The notion of cost is an important one, if we are to compare the performance of Sec.~\ref{baseline_approach} to that of Sec.~\ref{nested_approach}. We will compare costs in terms of a single QTG step, as depicted in Fig.~\ref{fig:unitofcost}. This step consists of a comparator and two adders, and its gate cost is $O(b^2)$ to leading order, where $b = \max(\lceil\log_2 c\rceil, \lceil\log_2 \left(\sum p_i\right)\rceil)$ is the maximal arithmetic register width \cite{QTG}. The cost of the entire QTG circuit on $n$ variables is hence $O(n)$ in this cost metric.
As described in \eqref{VanillaOPerator}, one Grover iteration applies the state preparation operator $A$ twice and the oracle and diffuser each once. In terms of gate complexity, the oracle can be implemented in $O(b)$ gates, and the diffuser in $O(n)$ gates. The gate complexity of $A$ per Grover iteration is $O(nb^2)$. In the unit of cost defined above, the oracle has cost $O(1)$, the diffuser $O(n/b^2)$ and the state preparation $O(n)$. For large enough instances (and therefore large enough $b$), the cost of the oracle and diffuser is negligible compared to that of the state preparation, and we will omit it in the following. A more detailed analysis of the cost of the oracles and diffuser is deferred to future work.

Using this reasoning, we will now set out to evaluate the cost of the approaches in Sec.~\ref{baseline_approach} and Sec.~\ref{nested_approach}. The baseline approach performed with $r$ Grover iterations per AA step, implements the operator described in \eqref{VanillaOPerator} where $A = QTG_n$ (the QTG circuit on all variables), $D = I-2\ket{0}\bra{0}$ and $O_{\text{global}}$ is the oracle defined in \eqref{oracles}. This implies that the cost of the baseline approach on a problem with $n$ items is
\begin{eqnarray}
    c_{\text{base}}(n, r) = n(2r+1).
\end{eqnarray} 
per AA step.
Consequently, as described thoroughly in Alg.~\ref{NestedGAS_pseudocode}, the operator implemented by the IIF with $r_\text{in}$ Grover iterations per AA step is
\begin{eqnarray}
    \label{inneroperator}
    X(r_\text{in}, y, k) = \left(-A D A^\dagger O_{\text{IIF}}(y)\right)^{r_{\text{in}}}A,
\end{eqnarray}
with $A = QTG_k$ (the QTG circuit on the first $k$ variables), $D = I-2\ket{0}\bra{0}$ and $O_{\text{IIF}}$ defined in \eqref{oracles}. Following an IIF routine with $r_\text{in}$ Grover iterations, the operator for the global GAS with $\tilde{r}$ Grover iterations per AA step is
\begin{equation}
    \label{outeroperator}
    \Bigl(-B X(r_\text{in}, y, k)\, D \bigl(B X(r_\text{in}, y, k)\bigr)^\dagger O_{\text{global}}(y)\Bigr)^{\tilde{r}}\; B\, X(r_\text{in}, y, k)
\end{equation}
Both \eqref{inneroperator} and \eqref{outeroperator} together form the cost for the nested approach. Given some depth $k$ at which the IIF is performed, $r_\text{in}$ Grover iterations in Alg.~\ref{IIF_pseudocode}, $\tilde{r}$ Grover iterations for the Global GAS and $n$ items, the cost of the nested approach per improvement step of the incumbent solution is
\begin{multline}
    \label{nested_cost}
    c_{\text{nested}}(n, k, r_\text{in}, \tilde{r}) = k+2kr_\text{in} + (n-k) \\
    + 2\tilde{r}\Bigl[(n-k)+2kr_\text{in}+k\Bigr] = (2\tilde{r}+1)(n+2r_\text{in}k).
\end{multline}
It is helpful to define a relative cost between both approaches over the course of a full run of the nested GAS. To ensure a symmetric comparison, we use a logarithmic scale. Let $C_{\text{base}}$ and $C_{\text{nested}}$ denote the total accumulated costs of the baseline and nested approaches respectively, where
\begin{eqnarray}
    C_{\text{base}} = \sum_{j=1}^{J} n(2r^{(j)}+1), \quad C_{\text{nested}} = \sum_{j=1}^{J'} c_{\text{nested}}^{(j)} + C_{\text{IIF}}^{(j)},
\end{eqnarray}
with $J$ and $J'$ the number of improvement steps taken by each approach before the termination condition is met, $r^{(j)}$ the number of Grover iterations in step $j$, $c_{\text{nested}}^{(j)}$ the per-step cost from~\eqref{nested_cost} and $C_{\text{IIF}}^{(j)}$ the IIF cost returned by Alg.~\ref{IIF_pseudocode}. The counter $C$ in Alg.~\ref{NestedGAS_pseudocode} tracks $C_{\text{nested}}$ during execution. The relative cost is then
\begin{eqnarray}
    c_{\text{rel}} = \log_2\left(\frac{C_{\text{base}}}{C_{\text{nested}}}\right).
\end{eqnarray}

\subsection{Termination criteria}\label{termination_criteria}
The termination for both the IIF in Alg.~\ref{IIF_pseudocode} and the global GAS in Alg.~\ref{NestedGAS_pseudocode} can be defined in various ways. The termination criterion for the global GAS depends on the accumulated cost $C$, which is tracked as a running counter in Alg.~\ref{NestedGAS_pseudocode}. After each iteration, $C$ is incremented by both the IIF cost $C_{\text{IIF}}$ returned by Alg.~\ref{IIF_pseudocode} and the outer GAS step cost $(2r+1)(n+2r_{\text{in}}k)$ as derived in \eqref{nested_cost}. Given a budget $\mathcal{B}$, the termination criterion reads:
\begin{eqnarray}\label{t_GAS}
    t_{\text{GAS}}(C) = \begin{cases}
    \text{True}, & \text{if } C \geq \mathcal{B} \\
    \text{False}, & \text{else.}
    \end{cases}
\end{eqnarray}
As shown in Alg.~\ref{IIF_pseudocode}, the termination criterion for the IIF is of a different nature. This criterion serves to validate whether the created superposition is beneficial. Generally, different approaches to this can be constructed. In this paper we chose to use a criterion that only terminates the IIF if a set number of measurements all landed in the marked subspace. Given some limit of measurements $L$ the simple criterion then reads:
\begin{eqnarray}
    \label{IIF_termination}
    t_{\text{IIF}}(s, L) = \begin{cases}
    \text{True}, & \text{if } s = L \\
    \text{False}, & \text{else,}
    \end{cases}
\end{eqnarray}
where $s$ is the number of successful measurements of a state in the marked subspace out of $L$ total measurements.
Looking at this sampling as a binomial process, the probability of measuring the marked subspace is directly related to the marked amplitudes after the IIF:
\begin{eqnarray}
    p = \sum_{x \in \text{marked}} |\bra{x}X(r_\text{in}, y, k)\ket{0}|.
\end{eqnarray}
The goal of any such termination as mentioned in \eqref{IIF_termination} is to maximize the value of $p$ with high confidence, while not making the cost of running the IIF too high. The Clopper-Pearson interval can be employed to set bounds for $p$, within some $p \in \left[p_\text{lower}, p_\text{upper}\right]$ \cite{Clopper_Pearson}, at a confidence of $1-\alpha$ where $\alpha$ is the significance level. Generally, one can write the bounds as:
\begin{eqnarray}
    B(\frac{\alpha}{2}; s, t-s+1) < p < B(1-\frac{\alpha}{2}; s+1, t-s),
\end{eqnarray}  
with $B$ the beta-distribution. Under the termination condition in \eqref{IIF_termination} a special case arises in which $p$ is upper-bounded by $p_{\text{upper}}=1$. For $L=5$, which was the empirically chosen standard, we obtain Table~\ref{CP_bounds}. It can be seen that choosing $L=5$ produces a reasonable confidence of strong marked subspace amplification. For instance, when all five measurements yield a marked state, we can be $90\%$ confident that the true success probability $p$ lies between $0.549$ and $1$, and more than half of the total amplitude is in the marked subspace. 
\begin{table}[H]
\centering
\setlength{\tabcolsep}{16pt}
\renewcommand{\arraystretch}{1.3}
\begin{tabular}{|c|c|c|}
\hline
$1-\alpha$ & $p_{\text{lower}}$ & $p_{\text{upper}}$ \\
\hline
 $80\%$& $0.631$ & $1$ \\
\hline
 $90\%$& $0.549$ & $1$ \\
\hline
 $95\%$& $0.478$ & $1$ \\
\hline
 $99\%$& $0.347$ & $1$ \\
\hline
\end{tabular}
\vspace{4pt}
\caption{Clopper-Pearson confidence intervals for different significance levels with $L=5$ measurements at IIF-level.}
\label{CP_bounds}
\end{table}

%% file: ClassicalTracking.tex
While quantum computers offer larger and larger numbers of qubits \cite{IBMnumberQubits,numberQubitsoverview,GoogleErrorCode} and executing optimization problems of bigger sizes becomes tractable \cite{largeScaleHybrid,largeScaleEnergy}, execution on hardware is still limited by noise and decoherence \cite{current_hardware}. Noise-free simulation of quantum circuits remains costly, even after the development of methods like tensor networks \cite{MPS1, MPS2, MPS3} or the use of high performance computing \cite{HPC}.

The problem of not being able to simulate larger than trivial problem sizes can be circumvented for the particular algorithm being studied in this paper. It turns out, that both the QTG as well as Grover's search in general and GAS particular have the property of allowing the efficient tracking of amplitudes of some ``bitstrings of interest". 

Given a certain incumbent value $y$, one first uses a classical solver kit (we use gurobipy-12.0.3)~\cite{gurobi} to extract the relevant marked subspaces for the knapsack instance. In particular, we extract two sets,
\begin{align}
    &S_{\text{global}}(y) = \big\{x \in \{0,1\}^n \;\big|\; \sum_{i=0}^{n-1} p_i x_i > y \nonumber \\
    &\qquad\qquad\qquad\qquad \land\; \sum_{i=0}^{n-1} w_i x_i \leq c\big\}, \\
    &S_{\text{partial}}(y, k) = \big\{x \in \{0,1\}^k \;\big|\; y {-} \sum_{i=k}^{n-1} p_i < \sum_{i=0}^{k-1} p_i x_i \nonumber \\
    &\qquad\qquad\qquad\qquad \land\; \sum_{i=0}^{k-1} w_i x_i \leq c\big\}.
\end{align}
The set $S_{\text{partial}}$ contains all partial states of length $k$ that fulfill \eqref{partial_capa} and \eqref{partial_val}, while $S_{\text{global}}$ is the set of all states that fulfill \eqref{capacity_constraint} and \eqref{vanilla_value}. More crucially, $S_{\text{global}}$ are ultimately the states we are interested in sampling from (meaning we need to know their amplitudes) following Alg.~\ref{NestedGAS_pseudocode}. States that are not in $S_{\text{global}}$ simply have the amplitude $1 - \sum_{x \in S_{\text{global}}} |\alpha_x|^2$, but their exact distribution is not relevant for discarding them as an unmarked subspace measurement. In the following, we will elaborate on how to precisely obtain the amplitude distribution of the states in $S_{\text{global}}$ after the nested approach.

As described in Sec.~\ref{QTG}, a single step of the QTG for item $i$ with some reference assignment $x^{\text{ref}}_i$ applies a biased Hadamard $H'$ controlled on whether item $i$ fits in the remaining capacity. For a state $x$ with current amplitude $\alpha^x$, the amplitude after one QTG step on item $i$ is
\begin{eqnarray}
    \alpha^x \mapsto \alpha^x \cdot h_i(x_i),
\end{eqnarray}
where
\begin{eqnarray}
    h_i(x_i) = \begin{cases}
    \sqrt{\frac{b+1}{b+2}}, & \text{if } x_i = x^{\text{ref}}_i \text{ and } \sum_{j=0}^{i} w_j x_j \leq c, \\[6pt]
    \sqrt{\frac{1}{b+2}}, & \text{if } x_i \neq x^{\text{ref}}_i \text{ and } \sum_{j=0}^{i} w_j x_j \leq c, \\[6pt]
    1, & \text{if } \sum_{j=0}^{i-1} w_j x_j + w_i > c \text{ and } x_i = 0,
    \end{cases}
\end{eqnarray}
and $b \geq 0$ is the bias parameter ($b=0$ recovers the standard Hadamard). The case where $\sum_{j=0}^{i-1} w_j x_j + w_i > c \text{ and } x_i = 0$ corresponds to the situation where item $i$ does not fit the remaining capacity, and the amplitude cannot be split. To now calculate the partial amplitude of some state $x \in S_{\text{partial}}$ we calculate
\begin{eqnarray}
    \alpha^x_{1, ..., k} = \prod_{i=0}^{k-1} h_i(x_i).
\end{eqnarray}
Using this marked subspace, one can then perform the calculation tracking through the AA in the IIF, which reduces to a multiplicative factor for all marked states and hence all states in $S_{\text{partial}}$. The amplitudes after the AA with $r_{\text{in}}$ Grover iterations are 
\begin{eqnarray}
    \label{IIF_amplitude}
    \tilde{\alpha}^x_{1, ..., k} = \alpha^x_{1, ..., k} \cdot \sin\bigl((2r_{\text{in}}+1)\theta_k\bigr),
\end{eqnarray}
where 
\begin{eqnarray}
\theta_k = \arcsin\sqrt{\sum_{x \in S_{\text{partial}}} |\alpha^x_{1, ..., k}|^2}.
\end{eqnarray}
In the step after the IIF, the QTG  is applied on the remaining variables ($\mathrm{QTG}_{n-k}$). For each globally marked state $x' \in S_{\text{global}}(y)$, it is necessarily true that $x'_{0:k} \in S_{\text{partial}}$, since marked states w.r.t. \eqref{vanilla_value} are marked w.r.t. \eqref{partial_val}. The amplitude of $x'$ after the QTG on the remaining $n-k$ variables is then
\begin{eqnarray}
    \alpha^{x'}_{1, ..., n} = \tilde{\alpha}^{x'}_{1, ..., k} \cdot \prod_{i=k}^{n-1} h_i(x'_i).
\end{eqnarray}
In the final step these amplitudes need to be tracked through the final AA with $r$ Grover iterations, yielding a final amplitude of some state $x' \in S_{\text{global}}(y)$ as
\begin{eqnarray}
    \alpha^{x'} = \alpha^{x'}_{1, ..., n} \cdot \sin\bigl((2r+1)\theta_n\bigr),
\end{eqnarray}
where 
\begin{eqnarray}
    \theta_n = \arcsin\sqrt{\sum_{x \in S_{\text{global}}} |\alpha^x_{1, ..., n}|^2}.
\end{eqnarray}
With these amplitudes it is now possible to sample from the set of marked states, emulating the shot-measurement of a quantum computer. Tracking the amplitudes of the baseline approach from Sec.~\ref{baseline_approach} follows the same logic, but without the need for $S_{\text{partial}}$.

%% file: results.tex
\subsection{Instance Generator} \label{instance_generator}
All instances for numerical results have been generated by the code provided with \cite{PisingerCode}. The instances, that can be generated have the following parameters:
\begin{itemize}
    \item $n$: number of items
    \item $r$: range s.t. weights are sampled as $w_i \in \{1,\dots,r\}$
    \item type: correlation type between profit and weight
    \begin{itemize}
        \item (uncorrelated): $p_i \in \{1,\dots,r\}$ sampled independently w.r.t. $w_i$
        \item (weakly correlated): $p_i \in [w_i-r/10,\,w_i+r/10]$ sampled uniformly
        \item (strongly correlated): $p_i = w_i + 10$

    \end{itemize}
    \item $S$ tightness of knapsack
    \item $i$: instance index (used as random seed)
    \item $c$: knapsack capacity
    $c = \text{min}(\frac{i}{S+1}\sum_{j=1}^{n} w_j, r+1)$.
\end{itemize}
Since the total knapsack capacity is capped from below by at least $r+1$, duplicate instances can be generated. These have been filtered out for the purposes of this paper.
\subsection{A note on value-weight correlation}
As described in Sec.~\ref{instance_generator}, the type of correlation between the values and the weights of items in the problem can help characterize the instance. In particular, correlation can have some influence on the difficulty of the instance. While uncorrelated instances are generally believed to be easiest to solve, exact trends are not straight-forward to understand \cite{WhereAreTheHardKnapsackInstances}.

The results for the nested approach (Sec.~\ref{nested_approach}) have been compiled for uncorrelated and for weakly correlated instances. Consistently, the results for the weakly correlated instances do not show any significant difference to the baseline approach from Sec.~\ref{baseline_approach}, or even performed worse. For this reason, the results for the weakly correlated instances are omitted from this section and are instead shown in Appendix~\ref{weak_results}. The results for the uncorrelated instances are shown in the following sections.
There is some value in discussing why the nested approach performs so poorly for correlated instances. As described in Sec.~\ref{QTG} and Sec.~\ref{nested_approach}, the nested approach amplifies all partial subspace that fulfill \eqref{capacity_constraint} and \eqref{partial_val}. For correlated instances, we know that $p_i \in [w_i-r/10,\,w_i+r/10]$ from Sec.~\ref{instance_generator}. For small $r$, \eqref{partial_val} roughly translates to the condition:
\begin{eqnarray}
    y - \sum_{i=k}^{n-1} p_i \lesssim \sum_{i=0}^{k-1} w_i x_i.
\end{eqnarray}
Hence amplified states lie roughly within an effective capacity band of $\tilde{c}=y - \sum_{i=k}^{n-1} p_i \lesssim \sum_{i=0}^{k-1} w_i x_i \leq c$. For uncorrelated instances given some partial bitstring, the weights $\{w_1, ..., w_k\}$ are essentially distributed at random when the system is sorted by value. For correlated instances this is not the case and the first $k$ weights will be on average the largest. It is however much more difficult to create low total weight partial states from a high-weight subset of the problem, which means that not many states will be sorted out because their total weight will tend to be bigger than $\tilde{c}$. Hence almost no states will be in the unmarked subspace for the IIF and the nested approach will perform poorly.

\subsection{Capacity to weight ratios}\label{sec:capweight}
\begin{figure}[H]
    \centerline{\includegraphics[width=0.5\textwidth]{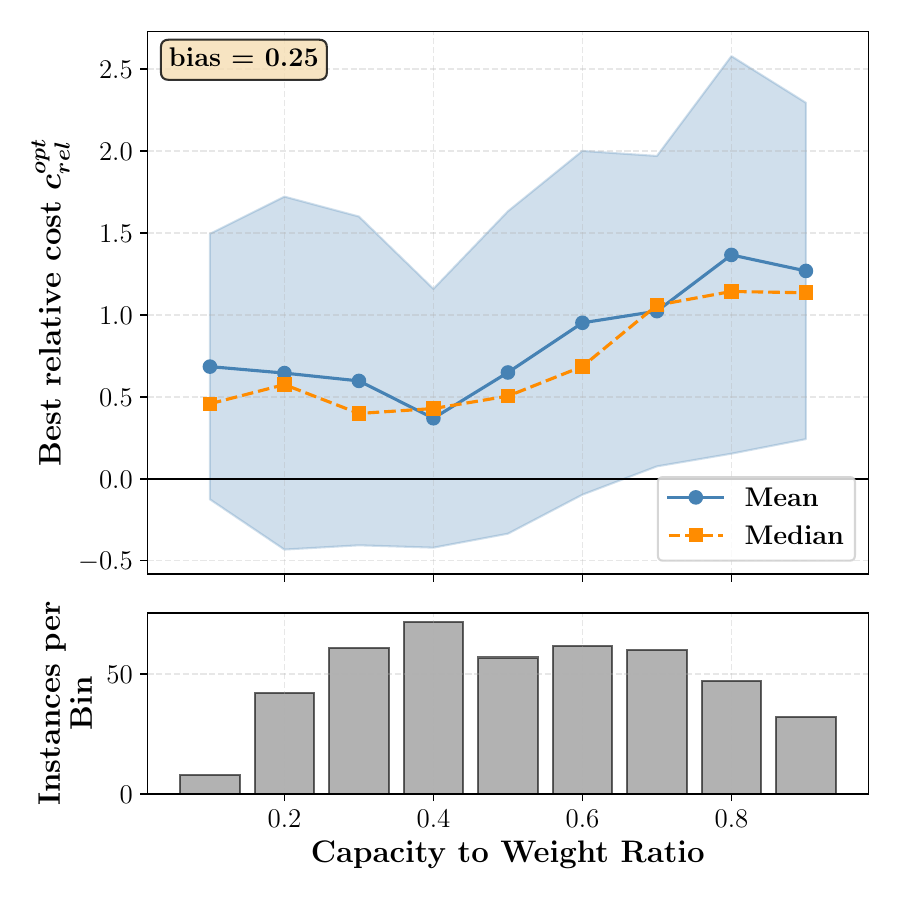}}
    \caption{The average of the relative cost differences for improving the greedy solution once, depending on the $\text{capweight}$ from \eqref{capweight}. For each instance, the depth $k$ is swept across different values and the lowest cost improvement is taken. The average and median are then shown with a $1\sigma$ envelope. Compiled over a set of instances spanning 10 to 35 items, with uncorrelated values and weights. The number of instances per $\text{capweight}$ bin is shown in the histogram below.}
    \label{result_capweight_unc}
\end{figure}

Another important characterization of a knapsack instance, is its capacity to weight ratio, defined as
\begin{eqnarray}
    \label{capweight}
    \text{capweight} = \frac{c}{\sum_{i=1}^{n} w_i}.
\end{eqnarray}
Generally speaking, $\text{capweight} > 1$ produces trivial instances (all items fit the knapsack), while $\text{capweight} < 1$ produces instances that can be hard to solve. A lower $\text{capweight}$ implies a more constrained instance. It can now be instructive to determine whether the nested approach is advantageous in certain regimes of $\text{capweight}$. To that end, we look at instances with different $\text{capweight}$ and determine the lowest cost improvement of the greedy solution for different depths $k$. Hence, each point contributing to the average taken in Fig.~\ref{result_capweight_unc} has cost:
\begin{eqnarray}
    c_{\text{rel}}^{\text{opt}} = \max_{k \in \{1, ..., n-1\}} c_{\text{rel}}.
\end{eqnarray}
Since, the optimum over all depths is considered, the result can in general not be used to determine whether the nested approach is advantageous for given instances, however it can be used to more broadly gauge regimes where the nested approach is most likely to perform well.

In particular, the results in Fig.~\ref{result_capweight_unc} suggest that the nested approach gains in effectiveness over the baseline approach as the knapsack becomes less constrained, i.e. for higher $\text{capweight}$. Instances with $0.6<\text{capweight}< 1$ appear to be the most promising, as the lowest possible cost to improve the greedy solution is significantly lower than for the baseline approach. The result was obtained by averaging over a set of instances with uncorrelated values and weights, spanning 10 to 35 items.

\subsection{Choosing an optimal depth on average}\label{sec:remaining_val}
Choosing an optimal depth $k$ for the nested approach is challenging generally. While previous work suggests that the optimal depth could be somewhere around $k = \frac{n}{2}$ \cite{NestedByGrover}, this is not necessarily the most advantageous depth for a general instance. More broadly speaking, if the weights and values in an instance are distributed very non-uniformly, more shallow/deeper depths could prove to be better. To quantify this, we introduce the $\text{RVTR}$, short for remaining value-threshold ratio:
\begin{eqnarray}
    \text{RVTR} = \frac{1}{y}\sum_{i=k+1}^{n} p_i,
\end{eqnarray} 
where $k$ is the depth and $y$ is the value of the incumbent. This ratio gives a better impression of depth in a way that it takes into account both the distribution of profits and the tightness of the incumbent solution.
\begin{figure}[H]
    \centerline{\includegraphics[width=0.5\textwidth]{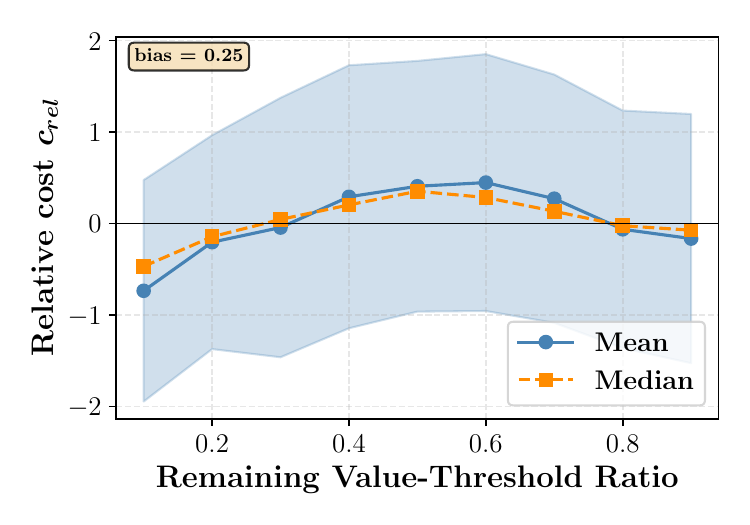}}
    \caption{The relative cost improvement of the nested approach is plotted over different values of the $\text{RVTR}$. The values for $\text{RVTR}$ are computed by both varying the incumbent and the depth $k$. The average, median and $1\sigma$ envelope are shown. Compiled over a set of instances spanning 10 to 35 items, with uncorrelated values and weights and capacity to weight ratios greater than 0.6. There appears to be a preferred region for $\text{RVTR}$ between $0.5$ and $0.6$ where the nested approach dominates the baseline approach.}
    \label{result_remaining_value_ratio}
\end{figure}
In Fig.~\ref{result_remaining_value_ratio}, we can see that there appears to be a clearly preferred range of $\text{RVTR}$ for the nested approach, which is around $0.5$ to $0.6$. In this range, the depth that results from a set $\text{RVTR}$ and some incumbent value $y$, shows on average the best cost improvement when improving the incumbent once, compared to the baseline approach. Notably, setting $k$ to be very early, or very late, produces results where the nested approach is worse, even on average. 

Since the $\text{RVTR}$ is classically efficiently computed given some depth $k$ we can employ this knowledge to consistently choose depths where the nested approach is likely to perform better than the baseline approach. Given some incumbent value $y$, we can compute the next depth on the uncorrelated instance as 
\begin{eqnarray}\label{optimal_depth}
     k  =  \argmin_{k \in \{1, ..., n-1\}} \lceil|\text{RVTR}(k, y) - 0.6|\rceil.
\end{eqnarray}
\subsection{Optimality Gap}
In Sec.~\ref{sec:capweight} and Sec.~\ref{sec:remaining_val}, we have identified specific regimes where we can expect the nested approach to outperform the baseline. In order to quantify to this advantage it is instructive to define several quantities. The first such quantity is the approximation ratio $\alpha$ defined as
\begin{eqnarray}
    \alpha(y) = \frac{y}{y^*},
\end{eqnarray} where $y$ is the value of the incumbent solution and $y^*$ is the value of the optimal solution. The approximation ratio of some solution $y$ quantifies its quality compared to the globally optimal solution. The second quantity of interest is the optimality gap, defined as
\begin{eqnarray}
    \gamma(y) = \frac{\alpha(y)-\alpha_{\text{greedy}}}{1 - \alpha_{\text{greedy}}},
\end{eqnarray}
where $\alpha_{\text{greedy}}$ is the approximation ratio of the greedy solution. The optimality gap quantifies how close a current incumbent $y$ is to the optimal solution, compared to the greedy solution. An optimality gap of $\gamma(y) = 0$ means that the incumbent solution is as good as the greedy solution, while $\gamma(y) = 1$ means that the incumbent solution is optimal. 

\begin{figure}[H]
    \centerline{\includegraphics[width=0.5\textwidth]{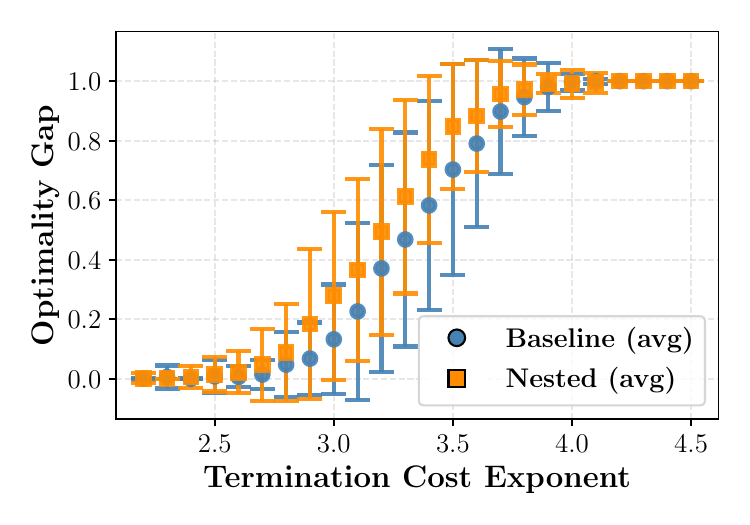}}
    \caption{The optimality with respect to the greedy solution is shown for the baseline and the nested approach, depending on the termination cost exponent $B$ defined in \eqref{budget}. The error bars indicate $1\sigma$ deviations from the mean. The results were compiled from uncorrelated instances, with $\text{capweight}>0.6$ and $C=10$. The optimal depth at each improvement step is calculated using \eqref{optimal_depth}. It can be seen that the nested approach outperforms the baseline approach.}
    \label{result_optgap_unc}
\end{figure}

In Fig.~\ref{result_optgap_unc} we can see the main result of this paper. Given some budget for the cost of both approaches, we show the optimality gap for the baseline and the nested approach. The budget is of the form
\begin{eqnarray}\label{budget}
    B = C \cdot n^t,
\end{eqnarray}
where $C$ is some constant, $n$ is the instance size and $t$ is the termination cost exponent. The budget is then used as a termination criterion as described in detail in Sec.~\ref{termination_criteria}. For the result in Fig.~\ref{result_optgap_unc}, we chose $C=10$. It can be observed, that given the same budget $B$, the nested approach consistently closes the optimality gap faster than the baseline approach. To avoid outliers due to sampling noise, the average comprises of datapoints that were each sampled $4$ different times. The result is obtained by averaging over a set of instances with uncorrelated values and weights, each with 40 items, and with $\text{capweight} > 0.6$.\\

Some notes on the result are in order. Firstly it is to be said that for these types of instances the greedy solution is already of high quality, with typically $\alpha_{\text{greedy}} > 0.95$. Secondly, the termination criteria described in Sec.~\ref{termination_criteria} omit the fact that it is also necessary to take the budget into account, when evaluating \eqref{IIF_termination}. However empirically, the cost contribution of the IIF is negligible compared to the global GAS. Lastly, the procedure of sampling each datapoint four times and averaging over the results is a way to mitigate the effect of sampling noise and see the advantageous trend on an average instance. However, it is to be noted that fundamentally, quantum computers are of course sampling based, so the average omits some of the intrinsic variance.

%% file: outlook.tex
In this paper, we have proposed and benchmarked a nested Amplitude Amplification scheme as an extension for GAS, applied to the binary knapsack problem. We showed that results can be obtained efficiently by classical simulation and used the QTG as a hot-starting procedure. As a result, we identified that the nested approach can be advantageous on average for uncorrelated instances in certain regimes of the capacity to weight ratio. We furthermore identified preferred depths to apply the nested approach.\\
In future work, a multitude of questions could be addressed. It should be instructive to look at tighter formulations for the partial amplification criterion in \eqref{partial_val}. Tighter conditions generally lead to more partial states being excluded from the marked subspace, improving the performance of nested approaches. These extensions however, always need to be studied in the light of the trade-off between the improved performance and the increased complexity of the circuit. Furthermore, the application of nested approaches to other optimization problems could be of interest. Lastly, an investigation into improving the nested approach for weakly correlated instances should be undertaken. Ideas include amplification criteria better suited to correlated instances or a more fitting reordering of the decision variables.\\

%% file: OracleConstruction.tex
Designing the oracle for binary knapsack problems proves to be straightforward, but should nevertheless be described here in some detail. As described in \eqref{oracles}, the oracles both simply compare whether the value encoded in the profit register is bigger than some classically computable value (either $T$ or $T - \sum_{i=k}^{n-1} p_i$) and if it is apply some phase kickback. Inspired by the approach in \cite{QTG}, the oracle is implemented in the following way. Let $(\tilde{p}_{log_{2} p}, ..., \tilde{p}_1)$ be the binary representation of the value stored in the quantum register. Furthermore, let $(\tilde{T}_{log_2 T}, ..., \tilde{T}_1)$ be the binary representation of the threshold value (or its modified version for $O_{\text{inner}}$). Checking whether the encoded value is bigger than the threshold value, can be done much more efficiently than controlling explicitly on all numbers bigger than $T$. Scanning from the most significant bit (MSB) to the least significant bit (LSB), there are three cases to consider at the first bit where $\tilde{p}$ and $\tilde{T}$ disagree. Let $j$ be the index of this bit, then:
\begin{enumerate}
    \item If $\tilde{p}[j] = 0$ and $\tilde{T}[j] = 1$, Then $p < T$,
    \item If $\tilde{p}[j] = 1$ and $\tilde{T}[j]$ = 0, Then $ p > T$,
    \item If no such $j$ is found, Then $p = T$.
\end{enumerate}
Since we only care about case 2, we control the phase operation on all qubits $j>k$ of the profit register, where $\tilde{T}_k = 0$, in such a way that $\tilde{p}_j = \tilde{T}_j  \forall j>k$ and $\tilde{p}_k = 1$. The cost of this oracle is of the order of $O(\text{log}_2T)$. The gate that is applied if the comparison is successful is an X-gate on a prepared flag register in initial state $\ket{-}$

%% file: Weakly_Correlated_Results.tex
\begin{figure}[H]
    \centerline{\includegraphics[width=0.5\textwidth]{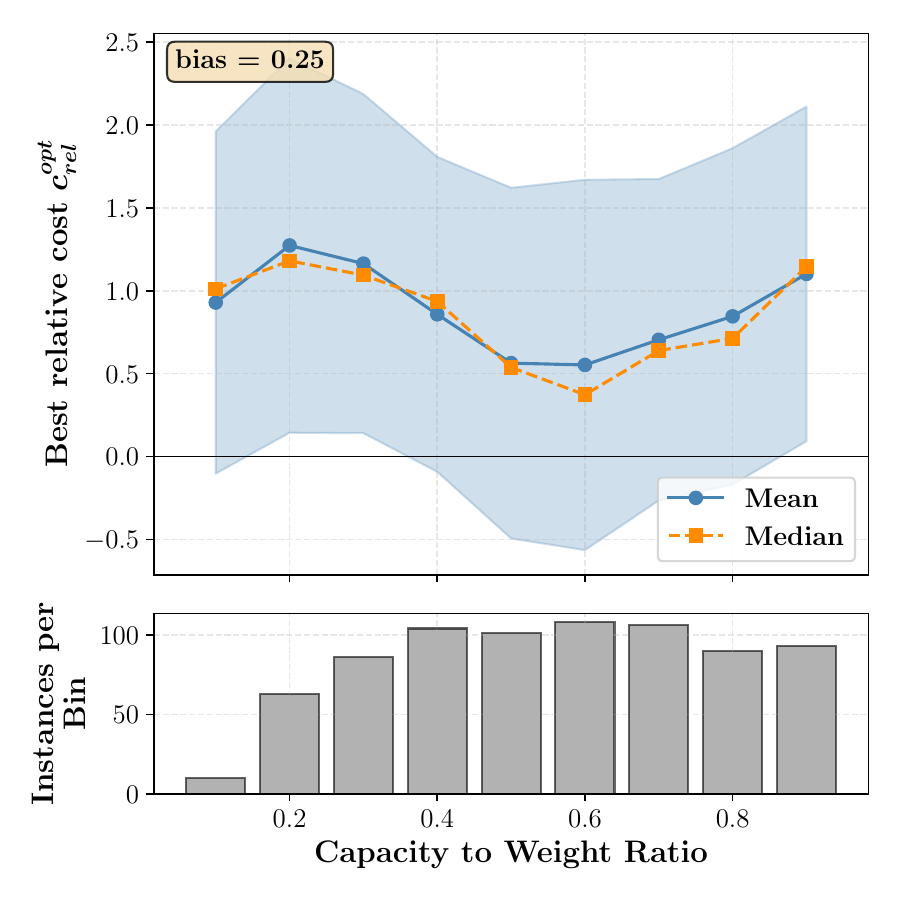}}
    \caption{The average of the relative cost differences for improving the greedy solution once, depending on the $\text{capweight}$ from \eqref{capweight}. For each instance, the depth $k$ is swept across different values and the lowest cost improvement is taken. The average and median are then shown with a $1\sigma$ envelope. Compiled over a set of instances spanning 10 to 35 items, with weakly correlated values and weights. The number of instances per $\text{capweight}$ bin is shown in the histogram below.}
    \label{result_capweight_weak}
\end{figure}

In this section, the same results as in Sec.~\ref{results} are shown for weakly correlated instances. \\
The result for weakly correlated instances when looking at the best possible cost improvement across the different depths, $c_{\text{rel}}^{\text{opt}}$, is shown in Fig.~\ref{result_capweight_weak}. One can clearly see that the result differ significantly from the uncorrelated ones. Particularly striking is the fact that for uncorrelated instances, higher $\text{capweight}$ appears to be an advantageous regime, whereas for the weakly correlated instances here, low $\text{capweight}$ appear better. Instances with $0.4 < \text{capweight}$ seem to be the most promising applications of the nested approach. \\

\begin{figure}[H]
    \centerline{\includegraphics[width=0.5\textwidth]{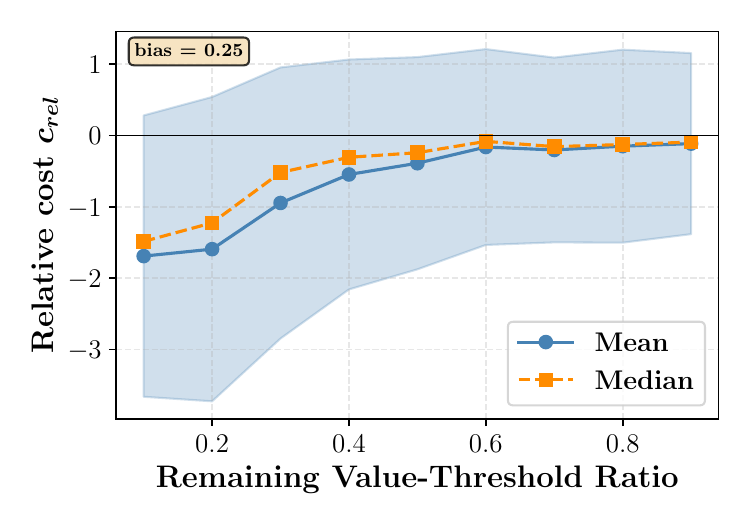}}
    \caption{The relative cost improvement of the nested approach is plotted over different values of the $\text{RVTR}$. The values for $\text{RVTR}$ are computed by both varying the incumbent and the depth $k$. The average, median and $1\sigma$ envelope are shown. Compiled over a set of instances spanning 10 to 35 items, with weakly correlated values and weights and capacity to weight ratios less than 0.4. The nested approach performs poorly compared to the baseline approach generally.}
    \label{result_remainingval_weak}
\end{figure}
Following the knowledge gained from Fig.~\ref{result_capweight_weak}, the study on the ideal depth will also be limited to such instances where $\text{capweight} < 0.4$. With such a limitation, Fig.~\ref{result_remainingval_weak} shows the results for weakly correlated instances of sizes between 10 and 35 items. The results show that the nested approach performs poorly compared to the baseline approach, even in the most promising regime of $\text{RVTR}$. This is in stark contrast to the uncorrelated instances, where a clear preferred region for $\text{RVTR}$ could be identified. The results suggest that for weakly correlated instances, the nested approach is generally not advantageous compared to the baseline approach. Since no optimal depth can be computed on average, we omit the results for the optimality gap.